\documentclass[conference]{IEEEtran}
\IEEEoverridecommandlockouts
\usepackage{cite}
\usepackage{amsmath,amssymb,amsfonts}
\usepackage{algorithmic}
\usepackage{graphicx}
\usepackage{textcomp}
\usepackage{xcolor}

\usepackage[colorlinks]{hyperref}
\hypersetup{colorlinks=false}
\definecolor{xiaolong}{RGB}{181, 97, 2}

\usepackage{cleveref}

\usepackage{booktabs} 

\renewcommand{\IEEEbibitemsep}{0pt plus 2pt}
\makeatletter
\IEEEtriggercmd{\reset@font\normalfont\footnotesize}

\makeatother
\IEEEtriggeratref{1}

\usepackage[boxed,ruled]{algorithm2e}
\usepackage{multirow}
\usepackage{makecell}
\usepackage{wrapfig}
\usepackage{setspace}

\usepackage[left=1.7cm,right=1.7cm,top=1.9cm,bottom=1.9cm]{geometry}

\renewcommand\baselinestretch{0.99}

\begin{document}

\title{\LARGE A SOT-MRAM-based Processing-In-Memory Engine for Highly Compressed DNN Implementation}

\author{\IEEEauthorblockN{Geng Yuan\text{$^\dagger$}}
\IEEEauthorblockA{\fontsize{8}{10.8}\textit{Northeastern University} \\
Boston, USA \\
yuan.geng@husky.neu.edu}
\and
\IEEEauthorblockN{Xiaolong Ma\text{$^\dagger$}}
\IEEEauthorblockA{\fontsize{8}{10.8}\textit{Northeastern University} \\
Boston, USA \\
ma.xiaol@husky.neu.edu}
\and
\IEEEauthorblockN{Sheng Lin}
\IEEEauthorblockA{\fontsize{8}{10.8}\textit{Northeastern University} \\
Boston, USA \\
lin.sheng@husky.neu.edu}
\and
\IEEEauthorblockN{Zhengang Li}
\IEEEauthorblockA{\fontsize{8}{10.8}\textit{Northeastern University} \\
Boston, USA \\
li.zhen@husky.neu.edu}
\and
\IEEEauthorblockN{Caiwen Ding}
\IEEEauthorblockA{\fontsize{8}{10.8}\textit{University of Connecticut} \\
Storrs, CT, USA \\
caiwen.ding@uconn.edu}
}

\maketitle

\def\footnoterule{\relax%
  \kern-5pt
  \hbox to \columnwidth{\hfill\vrule width 1\columnwidth height 0.4pt\hfill}
  \kern4.6pt}
\makeatother
\newcommand\blfootnote[1]{%
  \begingroup
  \renewcommand\thefootnote{}\footnote{#1}%
  \addtocounter{footnote}{-1}%
  \endgroup
}

\blfootnote{\hspace{-3.5mm}$^\dagger$These authors contributed equally.}

\begin{abstract}
The computing wall and data movement challenges of deep neural networks (DNNs) have exposed the limitations of conventional CMOS-based DNN accelerators.
Furthermore, the deep structure and large model size will make DNNs prohibitive to embedded systems and IoT devices, where low power consumption are required.
To address these challenges, 
spin orbit torque magnetic random-access memory (SOT-MRAM) and SOT-MRAM based Processing-In-Memory (PIM) engines have been used to reduce the power consumption of DNNs since SOT-MRAM has the characteristic of near-zero standby power, high density, none-volatile.
However, the drawbacks of SOT-MRAM based PIM engines such as high writing latency and requiring low bit-width data decrease its popularity as a favorable energy efficient DNN accelerator.
To mitigate these drawbacks, we propose an ultra energy efficient framework by using model compression techniques including weight pruning and quantization from the software level considering the architecture of SOT-MRAM PIM. And we incorporate the alternating direction method of multipliers (ADMM) into the training phase to further guarantee the solution feasibility and satisfy SOT-MRAM hardware constraints. Thus, the footprint and power consumption of SOT-MRAM PIM can be reduced, while increasing the overall system throughput at the meantime, making our proposed ADMM-based SOT-MRAM PIM more energy efficiency and suitable for embedded systems or IoT devices.
Our experimental results show the accuracy and compression rate of our proposed framework is consistently outperforming the reference works, while the efficiency (area \& power) and throughput of SOT-MRAM PIM engine is significantly improved.
 
\end{abstract}

\section{Introduction}

Large-scale DNNs achieve significant improvement in many challenging problems, such as image classification~\cite{krizhevsky2012imagenet}, speech recognition~\cite{amodei2016deep} and natural language processing~\cite{collobert2008unified}. However, as the number of layers and the layer size are both expanding, the introduced intensive computation and storage have brought challenges to the traditional Von-Neumann architecture~\cite{yuan2019ultra}, such as computing wall, massive data movement and high power consumption~\cite{yuan2017memristor, ding2017c}. Furthermore, the deep structure and large model size will make DNNs prohibitive to embedded systems and IoT devices, where low power consumption are required.

To address these challenges, 
spin orbit torque magnetic random-access memory (SOT-MRAM) has been used to reduce the power consumption of DNNs since it has the characteristic of near-zero standby power, high density, and none-volatile~\cite{umesh2018survey}. Combined with the in-memory computing technique~\cite{ma2018area,wang2018towards}, the SOT-MRAM based Processing-in-memory (PIM) engine could perform arithmetical and logic computations in parallel. Therefore, the most intensive operation, matrix multiplication and accumulation (MAC) in both convolutional layers (CONV) and fully-connected layers (FC), can be implemented using bit-wise parallel AND, bit-count, bit-shift, etc.

Compared to SRAM and DRAM, SOT-MRAM has higher write latency and energy~\cite{angizi2018imce, roohi2019processing, umesh2018survey}, which decrease its popularity as a favorable energy efficient DNN accelerator. To mitigate these drawbacks, from the software level, weight quantization has been introduced to SOT-MRAM PIM. By using binarized weight representation, ~\cite{angizi2018imce} efficiently processed data within SOT-MRAM to greatly reduce power-hungry and omit long distance data communication. However, weight binarization will cause accuracy degradation, especially for large-scale DNNs. In reality, many scenarios require as high accuracy as possible, e.g., self-driving cars. Thus, binarized weight representation is not favorable.

In this work, we propose an ultra energy efficient framework by using model compression techniques~\cite{lin2019toward,ma2019resnet,ding2018structured,ma2019pconv,liu2019autoslim} including weight pruning and quantization from the software level considering the architecture of SOT-MRAM PIM. 
To further guarantee 
the solution feasibility and satisfy SOT-MRAM hardware constraints while providing high solution quality (no obvious accuracy degradation after model compression and after hardware mapping), we incorporate the alternating direction method of multipliers (ADMM) into the training phase. As a result, we can reduce the footprint and power consumption of SOT-MRAM PIM, and improve the overall system throughput, making our proposed ADMM-based SOT-MRAM PIM more energy efficiency and suitable for embedded systems or IoT devices.

In the following paper, we first illustrate how to connect ADMM to our model compression techniques in order to achieve deeply compressed models that are tailored to SOT-MRAM PIM designs.
Then we introduce the architecture and mechanism of SOT-MRAM PIM and how to map our compressed model onto it. Finally, we evaluate our proposed framework on different networks. The experimental results show the accuracy and compression rate of our framework is consistently outperforming the baseline works. And the efficiency (area \& power) and throughput of SOT-MRAM PIM engine can be significantly improved.

 
\section{A Unified and Systematic Model Compression Framework for Efficient SOT-MRAM Based PIM Engine Design}

In this section, we propose a unified and systematic framework for DNN model compression, which simultaneously and efficiently achieves DNN weight pruning and quantization. By re-forming the pruning and quantization problems into optimization problems, the proposed framework can solve the structured pruning and low-bit quantization problems iteratively and analytically by extending the powerful ADMM~\cite{boyd2011distributed} algorithm.

In the meantime, our structured pruning model has a unique (i.e., tiny and regular) spatial property which naturally fits into the SOT-MRAM based processing-in-memory engine utilization, thereby successfully bridges the gap between the large-scale DNN inference and the computing ability limited platforms. After mapping the compressed DNN model on SOT-MRAM based processing-in-memory engine, bit-wise convolution can be executed with very high efficiency consistently.

\subsection{DNN Weight Pruning using ADMM}
For an $N$-layer DNN of interest. The first $M$ layers are CONV layers and the rest are FC layers. The weights and biases of the $i$-th layer are respectively denoted by ${\bf{W}}_{i}$ and ${\bf{b}}_{i}$, and the loss function associated with the DNN is denoted by $f \big( \{{\bf{W}}_{i}\}_{i=1}^N, \{{\bf{b}}_{i} \}_{i=1}^N \big)$; see \cite{zhang2018systematic}. In this paper, $\{{\bf{W}}_{i}\}_{i=1}^N$ and $\{{\bf{b}}_{i} \}_{i=1}^N$ respectively characterize the collection of weights and biases from layer $1$ to layer $N$.

In this paper, our objective is to implement structured pruning on the DNN. In the following discussion we focus on the CONV layers because they have the highest computation requirements.
More specifically, we minimize the loss function subject to specific structured sparsity constraints on the weights in the CONV layers, i.e., 
\begin{equation}
\label{opt0}
\begin{aligned}
& \underset{ \{{\bf{W}}_{i}\},\{{\bf{b}}_{i} \}}{\text{minimize}}
& & f \big( \{{\bf{W}}_{i}\}_{i=1}^N, \{{\bf{b}}_{i} \}_{i=1}^N \big),
\\ & \text{subject to}
& & {\bf{W}}_{i}\in {\bf{S}}_{i}, \; i = 1, \ldots, M,
\end{aligned}
\end{equation}
where ${\bf{S}}_{i}$ is the set of ${\bf{W}}_{i}$ with desired ``structure". Next we introduce constraint sets corresponding to different types of structured sparsity to facilitate SOT-MRAM PIM engine implementation. 

The collection of weights in the $i$-th CONV layer is a four-dimensional tensor, i.e., ${\bf{W}}_{i} \in R^{F_i \times C_i \times H_i \times W_i}$, where $F_i, C_i, H_i$, and $W_i$ are respectively the number of filters, the number of channels in a filter, the height of the filter, and the width of the filter, in layer $i$.

{\emph{\textbf{Filter-wise structured sparsity}}}:
When we train a DNN with sparsity at the filter level, the constraint on the weights in the $i$-th CONV layer is given by ${\bf{W}}_{i}\in {\bf{S}}_{i}
:=
\{{\bf{X}}\mid \text{the number of nonzero filters in}$ ${\bf{X}} ~\text{is less than or equal to}$ $\alpha_i \}.$ Here, nonzero filter means that the filter contains some nonzero weight.

{\emph{\textbf{Channel-wise structured sparsity}}}: When we train a DNN with sparsity at the channel level, the constraint on the weights in the $i$-th CONV layer is given by
${\bf{W}}_{i} \in {\bf{S}}_{i} :=  \{{\bf{X}}\mid \text{the number of nonzero }$ $\text{channels in}$ ${\bf{X}}~ \text{is less than or equal to}$ $\beta_{i}  \}.$
Here, we call the $c$-th channel nonzero if $({\bf{X}})_{:,c,:,:}$ contains some nonzero element.

{\emph{\textbf{Kernel-wise structured sparsity}}}:
When we train a DNN with sparsity at the Kernel level, the constraint on the weights in the $i$-th CONV layer is given by
${\bf{W}}_{i}\in {\bf{S}}_{i}
:=
 \{{\bf{X}}\mid \text{the number of }$ $\text{nonzero vectors in}$ $\{{\bf{X}}_{f,c,:,:}\}_{f,c=1}^{F_i,C_i}$ $ \text{is less than or equal to}~ \theta_i  \}.$

To solve the problem, we propose a systematic framework of dynamic ADMM regularization and masked mapping and retraining steps. We can guarantee solution feasibility (satisfying all constraints) and provide high solution quality through this integration.

\subsection{Solution to the DNN Pruning Problem}
Corresponding to every set ${\bf{S}}_{i}$, $i = 1, \ldots, M$ we define the indicator function
\[
g_{i}({\bf{W}}_{i})=
\begin{cases}
 0 & \text { if } {\bf{W}}_{i}\in {\bf{S}}_{i}, \\ 
 +\infty & \text { otherwise.}
\end{cases}
\]
Furthermore, we incorporate auxiliary variables ${\bf{Z}}_{i}$, $i = 1, \ldots, M$.
The original problem (\ref{opt0}) is then equivalent to 
\begin{equation}
\label{admm_form}
\begin{aligned}
& \underset{ \{{\bf{W}}_{i}\},\{{\bf{b}}_{i} \}}{\text{minimize}}
& & f \big( \{{\bf{W}}_{i} \}_{i=1}^N, \{{\bf{b}}_{i} \}_{i=1}^N \big)+\sum_{i=1}^{M} g_{i}({\bf{Z}}_{i}),
\\ & \text{subject to}
& & {\bf{W}}_{i}={\bf{Z}}_{i}, \; i = 1, \ldots, M.
\end{aligned}
\end{equation}

By adopting augmented Lagrangian \cite{boyd2011} on~\eqref{admm_form}, the ADMM regularization decomposes problem (\ref{admm_form}) into \emph{two} subproblems, and solves them iteratively until convergence. \\
\emph{\textbf{The first subproblem}} is
\begin{equation}
\label{subproblem_1}
 \underset{ \{{\bf{W}}_{i}\},\{{\bf{b}}_{i} \}}{\text{minimize}}
\ \ \ f \big( \{{\bf{W}}_{i} \}_{i=1}^N, \{{\bf{b}}_{i} \}_{i=1}^N \big)+\sum_{i=1}^{M} \frac{\rho_{i}}{2}  \| {\bf{W}}_{i}-{\bf{Z}}_{i}^{k}+{\bf{U}}_{i}^{k} \|_{F}^{2}, \\
\end{equation}
where ${\bf{U}}_{i}^{k}:={\bf{U}}_{i}^{k-1}+{\bf{W}}_{i}^{k}-{\bf{Z}}_{i}^{k}$.
The first term in the objective function of (\ref{subproblem_1}) is the differentiable loss function of the DNN, and the second term is a quadratic regularization term of the ${\bf{W}}_{i}$'s, which is differentiable and convex. As a result (\ref{subproblem_1}) can be solved by standard SGD. Although we cannot guarantee the global optimality, it is due to the non-convexity of the DNN loss function rather than the quadratic term enrolled by our method. Please note that this subproblem and solution are the same for all types of structured sparsities.\\
\emph{\textbf{The Second subproblem}} is
\begin{equation}
 \underset{ \{{\bf{Z}}_{i} \}}{\text{minimize}}
\ \ \ \sum_{i=1}^{M} g_{i}({\bf{Z}}_{i})+\sum_{i=1}^{M} \frac{\rho_{i}}{2} \| {\bf{W}}_{i}^{k+1}-{\bf{Z}}_{i}+{\bf{U}}_{i}^{k} \|_{F}^{2}. \\
\end{equation}
Note that $g_{i}(\cdot)$ is the indicator function of ${\bf{S}}_{i}$, thus this subproblem can be solved analytically and optimally \cite{boyd2011}. For $i = 1, \ldots, M$, the optimal solution is the Euclidean projection of ${\bf{W}}_{i}^{k+1}+{\bf{U}}_{i}^{k}$ onto ${\bf{S}}_{i}$.
The set ${\bf{S}}_{i}$ is different when we apply different types of structured sparsity, and the Euclidean projections will be described next.

Solving the second subproblem for different structured sparsities: For \textbf{filter-wise structured sparsity} constraints, we first calculate 
\[
\mathcal{R}_{f}=\| ({\bf{W}}_{i}^{k+1} +{\bf{U}}_{i}^{k})_{f,:,:,:} \|_F^2
\]
for $f = 1, \ldots, F_i$, where $\| \cdot \|_F$ denotes the Frobenius norm. We then keep $\alpha_i$ elements in $({\bf{W}}_{i}^{k+1} +{\bf{U}}_{i}^{k})_{f,:,:,:}$ corresponding to the $\alpha_i$ largest values in $\{\mathcal{R}_f\}_{f = 1}^{F_i}$ and set the rest to zero.

For \textbf{channel-wise structured sparsity}, we first calculate 
\[
\mathcal{R}_{c}=\| ({\bf{W}}_{i}^{k+1} +{\bf{U}}_{i}^{k})_{:,c,:,:} \|_F^2
\]
for $c= 1, \ldots, C_i$. We then keep $\beta_{i}$ elements in $({\bf{W}}_{i}^{k+1} +{\bf{U}}_{i}^{k})_{:,c,:,:}$ corresponding to the $\beta_{i}$ largest values in $\{\mathcal{R}_c\}_{c=1}^{C_i}$ and set the rest to zero.

For \textbf{kernel-wise structured sparsity}, we first calculate 
\[
\mathcal{R}_{f,c}=\| ({\bf{W}}_{i}^{k+1} + {\bf{U}}_{i}^{k})_{f,c,:,:} \|_F^2
\]
for $f= 1, \ldots, F_i$, $c= 1 \ldots, C_i$. We then keep $\theta_i$ elements in $({\bf{W}}_{i}^{k+1} + {\bf{U}}_{i}^{k})_{f,c,:,:}$ corresponding to the $\theta_i$ largest values in $\{\mathcal{R}_{f,c}\}_{f,c=1}^{F_i,C_i}$ and set the rest to zero.

\begin{figure} [t]
     \centering
     \includegraphics[width=0.8\columnwidth]{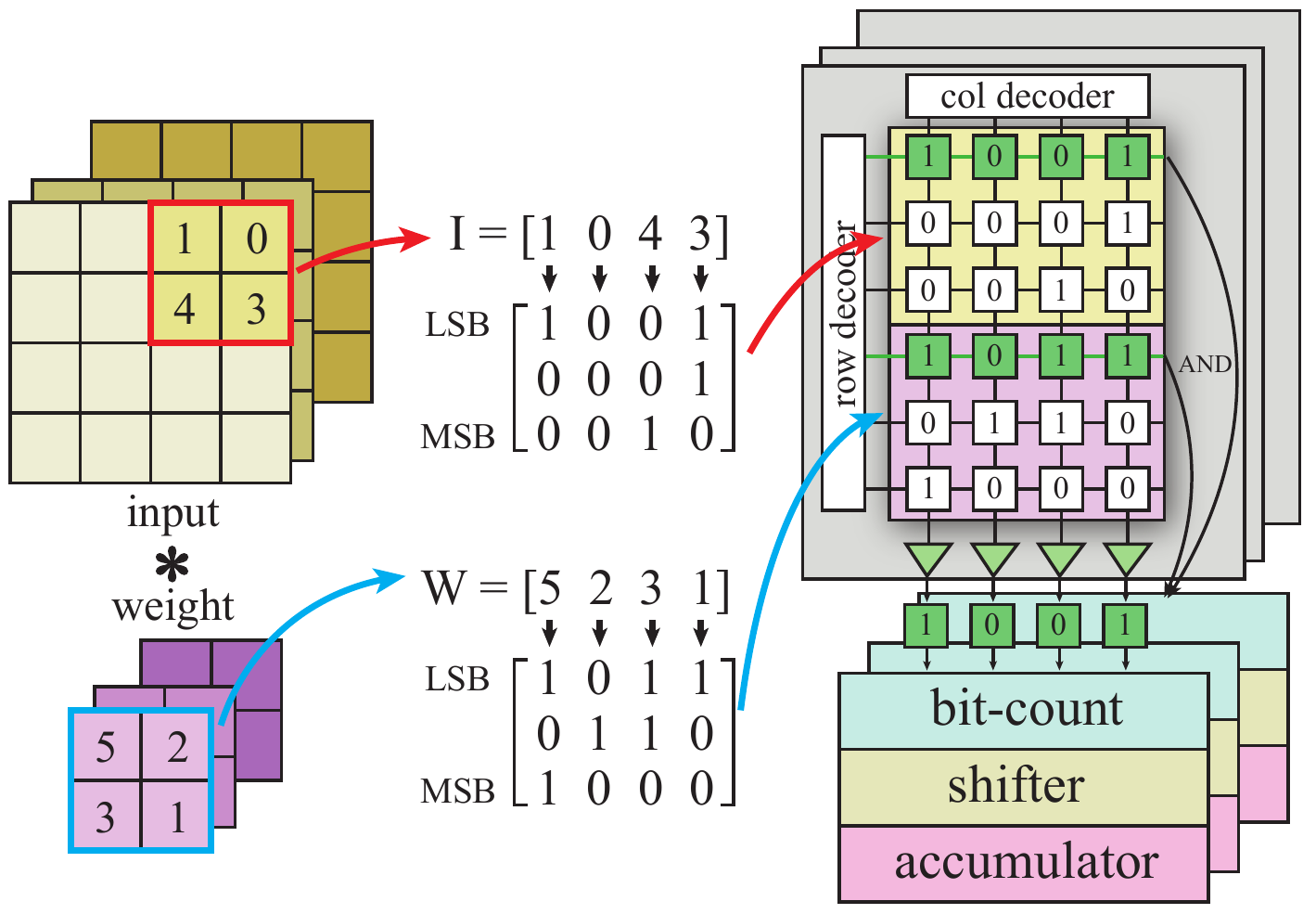}     
     \caption{Bit-wise convolution using SOT-MRAM based PIM engine}
     \label{fig:bit_count}
\end{figure}

\subsection{SOT-MRAM Processing-In-Memory Engine for DNN}

The main purpose of SOT-MRAM PIM engine is to convert and perform the MACs operations in convolutional layers using bit-wise convolution format. There are four main steps included in bit-wise convolutions: \textit{parallel AND operation, bitcount, bitshift and accumulation}. They can be formulated as Eqn.(\ref{bit-wise}). And $M$ and $N$ stands for the bit-length of inputs and weights respectively. The $c_m(I)$ represents the $m_{th}$-bit of all inputs in $I$, where the $c_n(W)$ contains the $n_{th}$-bit of all weights in $W$.

 Consider a CONV layer with 2$\times$2 kernel size, where $W$ contains 4 weights on a kernel and $I$ contains 4 current inputs covered by this kernel. We assume both weights and inputs are using 3-bit representation. From Figure~\ref{fig:bit_count} we can see that inputs and weights are mapped to two different sub-arrays. During the computation, two rows are selected from two sub-arrays each time, and the parallel AND results can be obtained from sense amplifiers~\cite{Fan2017}. Each row in one sub-array will conduct parallel AND operation with all the rows from the other sub-array. And every AND results will go through bit-count and shifter unit, then accumulated with other results to get a bit-wise convolution result~\cite{roohi2019processing}.

\begin{equation}
\label{bit-wise}
I*W=\sum_{m=0}^{M-1}\sum_{n=0}^{N-1}2^{m+n}bitcount(and(c_m(I), c_n(W)))
\end{equation}

\begin{figure} [t]
     \centering
     \includegraphics[width=0.9\columnwidth]{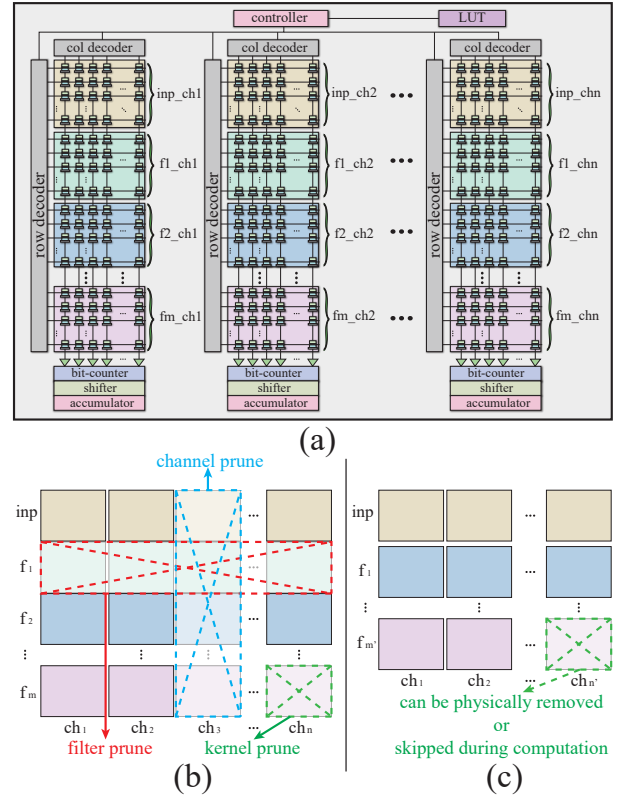}     \caption{Inputs and Weights mapping to the PIM Engine}
     \label{fig:sot_mapping}
\end{figure}

\subsection{Framework Mapping}

In our proposed framework, based on the architecture of SOT-MRAM-based PIM engine, we incorporate structured pruning and quantization using ADMM to effectively reduce the PIM engine's area and power. Quantization can be integrated in the same ADMM-based framework with different constraints. We omit the details of ADMM quantization due to space limit. Please note that our ADMM-based framework can achieve weight pruning and quantization simultaneously or separately.
Moreover, the overall throughput of SOT-MRAM based DNN computing system can be improved as well.
The SOT-MRAM-based PIM engine contains several processing elements (PEs). And each PE consists a column decoder, a row decoder, one computing set and multiple SOT-MRAM sub-arrays as shown in Figure~\ref{fig:sot_mapping}(1). It also shows how we map the inputs and weights to the PIM engine. In each PE, the inputs will be mapped on one sub-array, and every other sub-array will accommodate the different filters' weights from the same input channel. For example, the $PE1$ will compute the convolution of the inputs in $channel_1$ and all the weights in $channel_1$ from $filter_{1}$ to $filter_m$. And the number of columns and rows in each sub-array depends on the kernel size of the network and the bit-length of the inputs and weights respectively. All PEs are able to work parallelly and individually.

In Figure~\ref{fig:sot_mapping}(b), examples are given to show the corresponding structured pruning types that are used in our proposed framework and how it facilitates the reduction of PIM engine size. For the filter pruning, all the sub-arrays on the same row (i.e., storing the weights from the same filter) can be removed. Thus, the number of sub-arrays that are contained by each PE will be reduced. For the channel prune, since the pruned channels are no longer needed, the number of required PEs can be reduced without decreasing the throughput. Since each sub-array stores the weights from one channel of one filter, which is also considered as the weights from one convolution kernel, the kernel pruning will remove all the weights on a sub-array. 
By applying filter pruning and channel pruning, all the pruned sub-arrays or PEs can be physically removed directly. However, removing the sub-arrays by kernel pruning may cause an uneven size between different PEs. But since all the sub-arrays have same size and a sub-array is considered as a basic computing unit in bit-wise convolutions, the control overheads for addressing uneven sub-array numbers in PEs is ignorable. An alternative way is to use a look-up-table (LUT) to mark those pruned sub-arrays and skip them during computation instead of removing them physically.

Each pruning type has its own advantages. The filter pruning and channel pruning has propagation property. Because filter pruning (channel pruning) can not only remove the pruned weights but also removes the corresponding output channels (input channels) as well. By taking the advantage of that, the corresponding channels (filters) in next (previous) layer become redundant and can also be removed. The kernel pruning is especially tailored to the SOT-MRAM-based DNN computing system. Compared to filter and channel pruning, kernel pruning provides higher pruning flexibility, which means it is easier to maintain network accuracy under the same pruning rate. And none of them will incur complicated control logic.

The ADMM based quantization is also used in our proposed framework. In each weight sub-array, the number of rows equals to the bit-length that is used to represent the weights. The number of rows in each sub-array can be evenly reduced by quantizing the weights to fewer bits.


\begin{table}[t]
    \centering
    \caption{Structured pruning results on MNIST, CIFAR-10 and ImageNet using VGG-16 and ResNet-18/50 (RNT-18/50 in table). Accuracy results for ImageNet format as Top-1/Top5 accuracy.}
    \resizebox{0.48 \textwidth}{!}{
        \begin{tabular}{|c| l c c c|}
            \hline\hline
            & \multicolumn{1}{c}{\multirow{2}{*}{\makecell{\textbf{Method}}}} & \multirow{2}{*}{\makecell{\textbf{Base} \\ \textbf{Accuracy}}} &  \multirow{2}{*}{\makecell{\textbf{Prune} \\ \textbf{Accuracy}}} & \multirow{2}{*}{\makecell{\textbf{CONV} \\ \textbf{Comp. Rate}}} \\
            &&&& \multicolumn{1}{c|}{} \\\cline{1-5}
            \multicolumn{5}{|c|}{\textbf{MNIST (LeNet-5)}} \\ \hline
            \multirow{4}{*}{\textbf{\rotatebox[origin=c]{90}{}}} & Group Scissor~\cite{wang2017group} & 99.15\% & 99.14\% & 4.2$\times$  \\
            & \textbf{Our's} & \textbf{99.16\%} & \textbf{99.12\%} & \textbf{81.3$\times$}  \\
            \hline
            \multicolumn{5}{|c|}{\textbf{CIFAR-10}} \\ \hline
            \multirow{3}{*}{\textbf{\rotatebox[origin=c]{90}{RNT-18}}} & DCP~\cite{zhuang2018discrimination} & 88.9\% & 87.6\% & 2.0$\times$  \\
            & AMC~\cite{he2018amc} & 90.5\% & 90.2\% & 2.0$\times$  \\
            & \textbf{Our's} & \textbf{94.1\%} & \textbf{93.2\%} & \textbf{59.8$\times$}  \\
            \hline
            \multirow{3}{*}{\textbf{\rotatebox[origin=c]{90}{VGG-16}}} & Iterative Pruning~\cite{han2015learning,liu2018rethinking} &  92.5\%  & 92.2\% & 2.0$\times$  \\
            & 2PFPCE~\cite{min20182pfpce} & 92.9\% & 92.8\% & 4.0$\times$ \\
            & \textbf{Our's} & \textbf{93.7\%} & \textbf{93.3\%} & \textbf{44.8$\times$}   \\
            \hline
            \multicolumn{5}{|c|}{\textbf{ImageNet}} \\ \hline
            \multirow{3}{*}{\textbf{\rotatebox[origin=c]{90}{AlexNet}}} & Deep compression~\cite{han2015deep} & 57.2/82.2\% & 57.2/80.3\% & 2.7$\times$ \\
            & NeST~\cite{dai2017nest} & 57.2/82.2\% & 57.2/80.3\% & 4.2$\times$ \\
            & \textbf{Our's} & \textbf{57.4/82.4\%} & \textbf{57.3/82.2\%} & \textbf{5.2$\times$} \\
            \hline
            \multirow{3}{*}{\textbf{\rotatebox[origin=c]{90}{RNT-18}}} & Network Slimming~\cite{liu2017learning} & 68.9/88.7\% & 67.2/87.4\% & 1.4$\times$ \\
            & DCP~\cite{zhuang2018discrimination} & 69.6/88.9\% & 69.2/88.8\% & 3.3$\times$ \\
            & \textbf{Our's} & \textbf{69.9/89.1\%} & \textbf{69.1/88.4\%} & \textbf{3.0$\times$} \\
            \hline
            \multirow{3}{*}{\textbf{\rotatebox[origin=c]{90}{RNT-50}}} & Soft Filter Prune~\cite{he2018soft} & 76.1/92.8\% & 74.6/92.1\% & 1.7$\times$  \\
            & ThiNet~\cite{luo2017thinet} & 72.9/91.1\% & 68.4/88.3\% & 3.3$\times$ \\
            & \textbf{Our's} & \textbf{76.0/92.8\%} & \textbf{75.5/92.3\%} & \textbf{2.7$\times$} \\
            \hline

        \end{tabular}
    }
    \label{table:result_cifar}
\end{table}

\section{experimental results}
In our experiment, our generated compressed models are based on four widely used network structures, LeNet-5~\cite{lecun1998}, AlexNet~\cite{krizhevsky2012imagenet}, VGG-16~\cite{simonyan2014very} and ResNet-18/50~\cite{he2016deep}, and are trained on an eight NVIDIA RTX-2080Ti GPUs server using PyTorch~\cite{paszke2017pytorch}. 
For hardware results, we choose 32nm CMOS technology for the peripheral circuits. Cacti 7~\cite{balasubramonian2017cacti} is utilized to compute the energy and area of buffers and on-chip interconnects.
NVSim platform~\cite{Dong2012} with modified SOT-MRAM configuration is used to model the SOT-MRAM sub-arrays.
The power and area results of ADC are taken from~\cite{murmannadc}.

Several groups of experiments are performed, and we only show one result under each dataset and network, which achieves highest compression rate with minor accuracy degradation. Our results are based on 8-bit quantization, and we use combined pruning scheme (which means all three pruning types are used simultaneously). Table~\ref{table:result_cifar} shows our result on MNIST dataset using LeNet-5 can achieve 81.3$\times$ compression without accuracy degradation, which is 19.4$\times$ higher than Group Scissor~\cite{wang2017group}.
On CIFAR-10 dataset, our compression rates achieve 59.8$\times$ and 44.8$\times$ on ResNet-18 and VGG-16 networks with minor accuracy degradation. And on ImageNet, our compression rates for AlexNet, ResNet-18 and VGG-16 is 5.2$\times$, 3.0$\times$ and 2.7$\times$, respectively.

\begin{figure} [t]
     \centering
     \includegraphics[width=0.9\columnwidth]{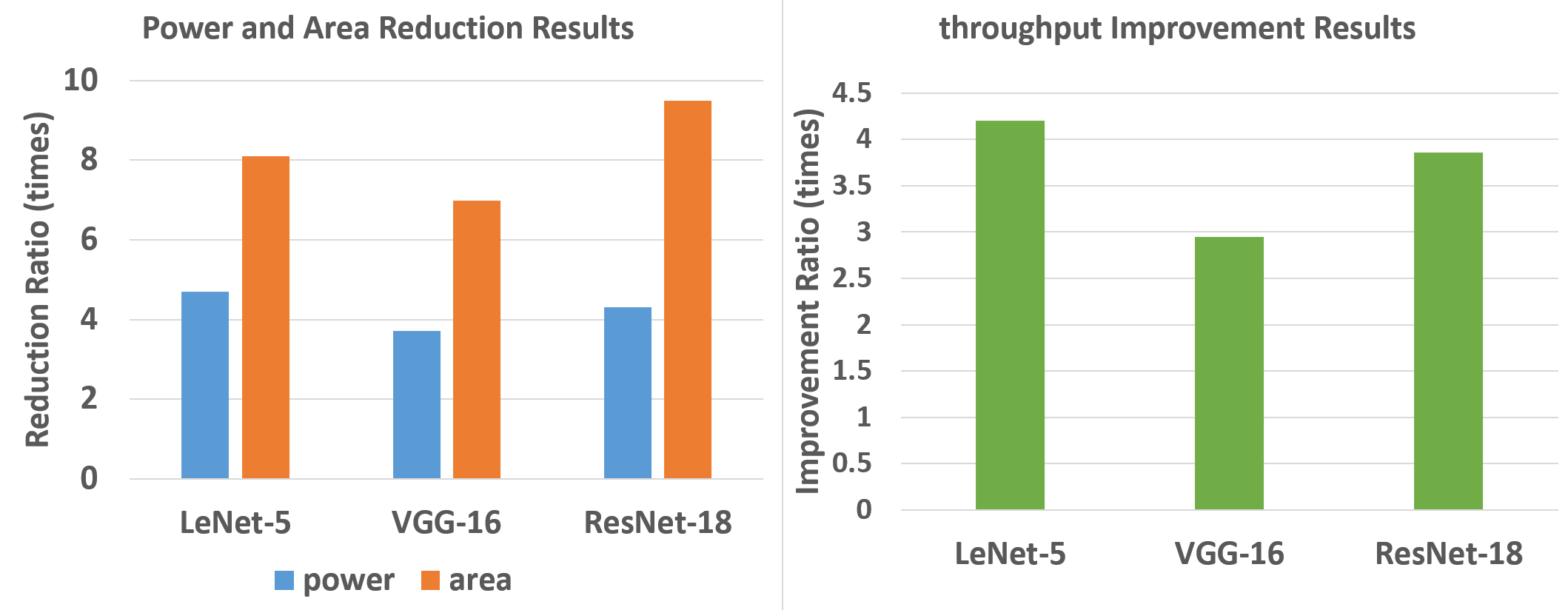}     \caption{Power/area reduction and throughput improvement over uncompressed models using MNIST and CIFAR-10 dataset.}
     \label{fig:power_area_throughput}
\end{figure}

By applying our framework, the power and area of SOT-MRAM PIM engine can be significantly reduced and the overall system throughput can also be improved comparing to uncompressed design, as shown in Figure~\ref{fig:power_area_throughput}. From our observation, channel pruning usually contributes more power and area saving than filter and kernel pruning, since it can remove entire PE with its peripheral circuits. On the other hand, the filter and kernel pruning can reduce the computing iterations between sub-arrays, which can improve the overall throughput.

\section{conclusion}

In this paper, we propose an ultra energy efficient framework by using model compression techniques including weight pruning and quantization at the algorithm level considering the architecture of SOT-MRAM PIM. 
And we incorporate ADMM into the training phase to further guarantee the solution feasibility and satisfy SOT-MRAM hardware constraints. The experimental results show the accuracy and pruning rate of our framework is consistently outperforming the baseline works. Consequently, the area and power consumption of SOT-MRAM PIM can be significantly reduced, while the overall system throughput is also improved dramatically.

\footnotesize{

}

\end{document}